\documentclass[aps,prx,reprint,superscriptaddress]{revtex4-1}

\usepackage{graphicx}
\usepackage{epstopdf}
\usepackage{epsfig}
\usepackage[utf8]{inputenc}
\usepackage[english]{babel}
\usepackage[colorlinks,urlcolor=blue,citecolor=blue]{hyperref}
\usepackage{amsmath}
\usepackage{amssymb}
\usepackage{dsfont}
\usepackage{verbatim}

\newcommand{\D}{\ensuremath{\mathcal{D}}}
\newcommand{\E}{\ensuremath{\mathcal{E}}}
\renewcommand{\H}{\ensuremath{\mathcal{H}}}
\renewcommand{\L}{\ensuremath{\mathcal{L}}}
\newcommand{\dd}{\ensuremath{\mathrm{d}}}
\newcommand{\dt}{\dd{}t}
\newcommand{\dW}{\dd{}W}
\newcommand{\Hint}{\ensuremath{H_\mathrm{int}}}
\newcommand{\avg}[1]{\ensuremath{\langle #1\rangle}}

\newcommand{\covU}{\ensuremath{\Gamma^\mathrm{u}}}
\newcommand{\covC}{\ensuremath{\Gamma^\mathrm{c}}}
\newcommand{\tr}{\ensuremath{\mathrm{tr}}}

\newcommand{\nbar}{\ensuremath{\bar{n}}}
\newcommand{\outpr}[2]{\ensuremath{| #1\rangle\langle #2 |}}

\newcommand{\SME}{stochastic master equation}

\newcommand{\eref}[1]{(\ref{#1})}
\newcommand{\eq}[1]{Eq. \eref{#1}}

\begin{document}


  \title{Adiabatic Elimination of Gaussian Subsystems from Quantum Dynamics under Continuous Measurement}

  \author{Ond\v{r}ej \v{C}ernot\'ik}
  \email{Ondrej.Cernotik@itp.uni-hannover.de}
  \affiliation{Institute for Theoretical Physics, Institute for Gravitational Physics
    (Albert Einstein Institute), Leibniz University Hannover, Callinstra\ss{}e 38,
    30167 Hannover, Germany}
  \author{Denis V. Vasilyev}
  \affiliation{Institute for Theoretical Physics, Institute for Gravitational Physics
    (Albert Einstein Institute), Leibniz University Hannover, Callinstra\ss{}e 38,
    30167 Hannover, Germany}
  \affiliation{Department of Physics, University of York, Heslington, York YO10 5DD, United Kingdom}
  \author{Klemens Hammerer}
  \affiliation{Institute for Theoretical Physics, Institute for Gravitational Physics
    (Albert Einstein Institute), Leibniz University Hannover, Callinstra\ss{}e 38,
    30167 Hannover, Germany}

  \date{\today}

  \begin{abstract}
    An ever broader range of physical platforms provides the possibility to study and engineer quantum dynamics under continuous measurements. In many experimental arrangements the system of interest is monitored by means of an ancillary device, whose sole purpose is to transduce the signal from the system to the measurement apparatus. Here, we present a method of adiabatic elimination when the transducer consists of an arbitrary number of bosonic modes with Gaussian dynamics while the measured object can be any quantum system. Crucially, our approach can cope with the highly relevant case of finite temperature of the transducer, which is not easily achieved with other methods. We show that this approach provides a significant improvement in the readout of superconducting qubits in circuit QED already for a few thermal excitations, and admits to adiabatically eliminate optomechanical transducers.
  \end{abstract}

  \maketitle


  \section{Introduction}

  Quantum limited, continuous measurements and measurement-based feedback \cite{Wiseman2010,Zhang2014} represent important concepts for fundamental studies of open quantum systems and the measurement process in quantum mechanics, and beyond that they are highly useful tools for applications in Quantum Information Processing. Since the first demonstration of quantum limited, continuous measurements with single ions \cite{Bushev2006} and atoms \cite{Kubanek2009} the concepts of quantum dynamics under continuous monitoring have gained great experimental relevance in recent years in the field of circuit QED \cite{Arakawa2015}. Here, measurement and feedback have been used for preparation of qubit states \cite{Johnson2012, Riste2012, Riste2012a} including preparation of entangled states \cite{Riste2013, Roch2014}, or for observing and stabilization of quantum trajectories \cite{Vijay2012, Murch2013, Weber2014}. 
  Only very recently cavity optomechanical systems \cite{Aspelmeyer2013} entered the parameter regime of quantum limited, continuous measurement \cite{Purdy2013} and feedback within the thermal decoherence time \cite{Wilson2014}, where the tools of continuous measurement theory will unfold their full strength.

\begin{figure}
    \centering
    \includegraphics[width=\linewidth]{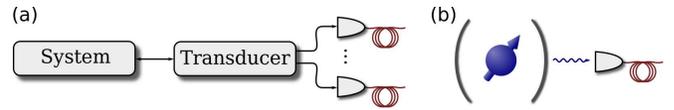}
    \caption{\label{fig.composite}
      (a) Schematic of the considered setup: A quantum system is monitored by coupling it to an ancillary system -- the transducer -- whose continuously emitted light field is detected.  The most simple example of such a setup---a qubit coupled to a cavity mode with monitored output---is shown in (b). The transducer can also be a much more complex device, e.g. an optomechanical converter of microwave to optical photons.
    }
  \end{figure}

  In a typical measurement scenario, the system of interest (e.g., a superconducting qubit or a mechanical oscillator)
  interacts with an ancillary system, such as a cavity mode, whose continuously emitted output field is then sent to a measurement device, cf. Fig.~\ref{fig.composite}. In the most simple case the ancilla is a single cavity mode transducing the signal photons emitted from e.g. a superconducting qubit to the measurement apparatus. Beyond that, the ancillary transducer can consist of a much more sophisticated subsystem, e.g. an arrangement for frequency conversion of photons from microwave to optics, which is subject to its own nontrivial dynamics, losses and sources of thermal noise. The precise dynamics of the auxiliary system is oftentimes irrelevant  and we are interested only in obtaining the equation of motion of the system alone.  Obtaining the reduced dynamics of the system is crucial for two reasons: Firstly,  auxiliary systems quickly make any numerical simulations intractable due to the increased Hilbert space dimension. This becomes especially troublesome when the transducer is thermally excited at finite temperatures. Secondly, in more complex setups, where the transducer is composed of  several coupled subsystems, the structure of the ancilla easily obscures the effect of the continuous measurement on the main system, and makes it difficult to design feedback protocols.

    The dynamical degrees of freedom of the transducer can be adiabatically eliminated from the dynamics if their evolution (e.g. the cavity decay) is fast on the time scale of the interaction with the system. In the context of stochastic quantum dynamics under continuous measurements perturbative techniques used for adiabatic elimination of fast degrees of freedom received significant attention in the theoretical literature in this field \cite{Doherty1999, Gough2007, Hutchison2009, Gambetta2008, Bouten2008, Bouten2008a, Lalumiere2010,Yang2012}. However, these methods become cumbersome or intractable when the ancillary system becomes large (i.e. consists itself of several subsystems) and/or thermally excited at finite temperature. Both of these cases are highly relevant for current experiments: On the one hand, thermal excitations typically cannot be neglected in the frequency domain of circuit quantum electrodynamics \cite{Arakawa2015}. On the other hand, optomechanical systems composed of several coupled mechanical, optical and microwave modes can be used as transducers in order to convert photons at vastly different length scales \cite{Tian2010,Stannigel2010,Taylor2011,Bochmann2013,Bagci2014,Andrews2014}.

  In this paper, we present a new method for adiabatic elimination in conditional dynamics which applies to transducers composed of an arbitrary number of bosonic modes whose dynamics is Gaussian \cite{Weedbrook2012, Edwards2005},  that is, the dynamics is generated by a quadratic Hamiltonian, linear jump (decay) operators, and the transducer is subject to continuous homodyne detection.  We use the fact that dynamics of such systems can be described using their first and second statistical moments---the
  mean values and the covariance matrix---to obtain a stochastic master equation for the system of interest only.
  The main advantage of this method is the possibility to eliminate a subsystem of arbitrary dimension coupled to thermal bathes
  which implies a broad range of applications for superconducting and optomechanical systems.

  We introduce the method in Sec. \ref{sec.elimination}, and illustrate the method with several examples in Sec. \ref{sec.examples}.
  Not only do we show that our method of adiabatic elimination outperforms more na\"ive approaches
  when dispersively reading out a qubit using a cavity with just a few thermal excitations,
  we also study measurement-induced entanglement generation in such a setting, generalizing the results of Ref. \cite{Hutchison2009}.
  Finally, in Sec. \ref{sec.conclusions} we conclude and  suggest other applications of our method in quantum control scenarios with optomechanical transducers  \cite{Tian2010,Stannigel2010,Taylor2011,Bochmann2013,Bagci2014,Andrews2014}.


  \section{Effective dynamics}\label{sec.elimination}
  \subsection{Main results}\label{ssec.results}

  The measurement scenario we have in mind is illustrated in Fig. \ref{fig.composite}.
  The system of interest (e.g., a qubit) couples to a transducer (e.g., a cavity mode) whose output fields are continuously measured in a homodyne detection.  The conditional dynamics of the overall system, including losses, noise, and the effect of continuous diffusive measurement, is described by the stochastic master equation
  \begin{equation}\label{eq.sme}
    \dd\rho = \L_S\rho\dt+\L_T\rho\dt+\L_\mathrm{int}\rho\dt+\sum_m\H[\lambda_m]\rho\dW_m.
  \end{equation}
  $\L_S$ is the system Liouvillian that contains, in general, some coherent dynamics given by a system Hamiltonian $H_S$ and some Lindblad operators decribing decoherence, but will be left unspecified for the moment.
  The Liouvillian for the transducer is
  \[
  \L_T\rho = -i[H_T,\rho]+\sum_i\D[j_i]\rho,
  \]
  where $H_T$ is the Hamiltonian and the Lindblad terms $\D[j_i]\rho = j_i\rho j_i^\dagger-\frac{1}{2}(j_i^\dagger j_i\rho+\rho j_i^\dagger j_i)$ describe decoherence and measurement channels. We further assume that the transducer is Gaussian, i.e. it consists of $N$ bosonic modes with Hamiltonian  $H_T$ and jump operators $j_i$ which are, respectively, quadratic and linear in canonical operators.  It is convenient to collect the canonical operators into the $2N$-dimensional vector $r = (q_1,p_1,\ldots,q_N,p_N)^T$
  with commutation relations $[r_i,r_j] = i\sigma_{ij}$ and
  $\sigma = \bigoplus_{i=1}^N\begin{pmatrix}0&1\\-1&0\end{pmatrix}$
  being the symplectic matrix.
  The transducer Hamiltonian and the jump operators can then be expressed as, respectively,
  \begin{align*}
  H_T &= \frac{1}{2}r^TRr, & j_i &= \xi_i^Tr,
  \end{align*}
   with a real symmetric matrix
  $R = R^T\in\mathbb{R}^{2N}\times\mathbb{R}^{2N}$ and complex vectors $\xi_i\in\mathbb{C}^{2N}$.
  Furthermore, we assume the interaction between system and transducer linear in the transducer operators
  \begin{align*}
  \L_\mathrm{int}\rho &= -i\epsilon[\Hint,\rho], & \Hint &= s^Tr.
  \end{align*}
  where $s$ is a $2N$-dimensional vector of Hermitian operators acting on the system $S$.
  We use the small parameter $\epsilon$ to remind us that the interaction is weak and can be treated perturbatively.
  Finally, for the transducer to be Gaussian, the measurement terms correspond to a homodyne detection,
  i.e.,
  \[
  \H[\lambda_m]\rho = (\lambda_m-\avg{\lambda_m})\rho+\rho(\lambda_m^\dagger-\avg{\lambda_m^\dagger}),
  \]
  and the measurement operators are linear in the canonical operators,
  \begin{align}\label{eq.cm_mm}
  \lambda_m = (c_m+im_m)^Tr,
  \end{align}
  with $c_m, m_m\in\mathbb{R}^{2N}$.
  The measurments are independent, $\dW_m\dW_n = \delta_{mn}\dt$ and for each measurement operator $\lambda_m$,
  there should be a corresponding Lindblad term $\D[\lambda_m]$ in the transducer Liouvillian $\L_T$.
  The measurements give rise to classical measurement currents that take the form
  \begin{equation}\label{eq.dI}
    \dd I_m = \avg{\lambda_m+\lambda_m^\dagger}\dt+\dW_m.
  \end{equation}

  To zeroth order in the coupling parameter $\epsilon$, the transducer dynamics is Gaussian
  which means that it can be fully described using the first and second statistical moments of the canonical operators,
  i.e., the mean values $x_i(t) = \avg{r_i(t)} = \tr\{\rho_T(t)r_i\}$,
  and the covariance matrix with an element $\covC_{ij}(t) = \avg{[r_i,r_j]_+(t)} -2x_i(t)x_j(t) =
  \tr\{\rho_T(t)[r_i,r_j]_+\}-2x_i(t)x_j(t)$;
  here $\rho_T=\tr_S\{\rho\}$ is the reduced density operator of the transducer (in zeroth order of $\epsilon$) and we use the superscript $\mathrm{c}$ to indicate
  that the moments are calculated with respect to the conditional state $\rho_T$, obeying the stochastic master equation
  \begin{align}\label{eq.condMEQ}
  \dd\rho_T = \L_T\rho_T\dt+\sum_m\H[\lambda_m]\rho_T\dW_m.
  \end{align}
  For the mean values and the covariance matrix this implies the following equations of motion
  \begin{subequations}
  \begin{eqnarray}
    \dd x &=& Ax\dt+\sum_m(\covC c_m-\sigma m_m)\dW_m,\label{eq.conditional_means}\\
    \dot{\Gamma}^\mathrm{c} &=& A\covC+\covC A^T+2N-\nonumber\\
      &&-2(\covC c_m-\sigma m_m)(\covC c_m-\sigma m_m)^T,\label{eq.Riccati}
  \end{eqnarray}
  \end{subequations}
  where
  \begin{subequations}\label{eq.defs}
  \begin{eqnarray}
    A &=& \sigma R-\frac{i}{2}\sigma\sum_i(\xi_i^\dagger\xi_i-\xi_i^T\xi_i^\ast),\\
    N &=& \frac{1}{2}\sigma\sum_i(\xi_i^\dagger\xi_i+\xi_i^T\xi_i^\ast)\sigma^T;
  \end{eqnarray}
  \end{subequations}
  see Appendix \ref{app.Gaussian} for details.

  Averaging Eq.~\eqref{eq.condMEQ} over the measurement record, we recover the deterministic master equation for the unconditional state
  \[
  \dot{\rho}_T^\mathrm{u} = \L_T\rho_T^\mathrm{u}
  \]
  (we use the superscript $\mathrm{u}$ to indicate that the state is unconditional)
  and the corresponding equations of motion for the first and second moments
  \begin{subequations}
  \begin{eqnarray}
    \dot{x}^\mathrm{u} &=& Ax^\mathrm{u},\\
    \dot{\Gamma}^\mathrm{u} &=& A\covU + \covU A^T+2N.\label{eq.Lyapunov}
  \end{eqnarray}
  \end{subequations}
   Note that in both cases, conditional and unconditional dynamics, the covariance matrix obeys a deterministic equation of motion of Ricatti or Lyapunov type, cf. Eqs.~\eqref{eq.Riccati} and \eqref{eq.Lyapunov} respectively, which can be solved efficiently.

  Our main goal is to derive a closed, effective equation of motion for the conditional state of the system $\rho_S=\tr_T\{\rho\}$ which is correct to leading order of $\epsilon$ based on the assumption that transducer dynamics $\L_T$ is fast on the time scale of the system-transducer interaction $\Hint$. Under this condition the state of the system will be given by $\rho = \rho_S\otimes\rho_T+O(\epsilon)$ where $\rho_T$ is the steady state solution of Eq.~\eqref{eq.condMEQ}. The strategy now is to determine equations of motion for the order-$\epsilon$ correction to this approximation, solve them formally, and substitute the solution into the equation of motion for $\rho_S$. In this way we can arrive at the closed, effective equation of motion for $\rho_S$, which will be of second order in $\epsilon$ in the deterministic  and of first order in the stochastic part, as we will see. In the following we will summarize the final result of this adiabatic elimination procedure. The derivation is given in the next section.

  So far we have left the system dynamics $\L_S$ unspecified. For the adiabatic elimination to work we will have to make assumption regarding $\L_S$ relative to $\Hint$ and  $\L_T$. We will consider two main regimes:

  (a)  The system dynamics is trivial, $\L_S = 0$. This can be fulfilled exactly in an interaction picture when the system operators $s_j$ in $\Hint$ happen to be constants of motion, and covers in particular the important case of a quantum nondemolition measurement. Else, $\L_S = 0$ can be fulfilled approximately if the time scales of $\L_S$ are much slower than those of $\Hint$. Under this assumption the effective equation of motion for the state of the system $\rho_S$ is found to be
  \begin{align}\label{eq.system_DC}
    \dd\rho_S &= \frac{1}{2}A_{ij}^{-1}\covU_{jk}[s_i,[s_k,\rho_S]]\dt+\frac{i}{2}A_{ij}^{-1}\sigma_{jk}[s_i,[s_k,\rho_S]_+]\dt
        \nonumber\\
      &\quad+\H[i\Lambda_m^Ts]\rho_S\dW_m,
  \end{align}
  where we have for the measurement term
  \begin{subequations}
  \begin{eqnarray}
    \Lambda_m &=& (\covC-i\sigma)Q^{-T}c_m+A^{-1}(\covC c_m-\sigma m_m), \\
    Q &=& A-2(\covC c_m-\sigma m_m)c_m^T.\label{eq.q}
  \end{eqnarray}
  \end{subequations}
  We remind the reader that $\Gamma^\mathrm{c(u)}$ refer to the covariance matrix of the transducer attained as steady state solutions of Eqs.~\eqref{eq.Riccati} and \eqref{eq.Lyapunov}, respectively. The coefficients $c_m$ and $m_m$ are given in \eqref{eq.cm_mm}, and the matrix $A$ in \eqref{eq.defs}.  In the last equation and in following equations we use the Einstein summation convention.

  As expected, the deterministic part of the stochastic master equation (first line in Eq.~\eqref{eq.system_DC}) depends only on the unconditional state of the transducer through its covariance matrix $\Gamma^\mathrm{u}$. Note that the stochastic term does depend on the conditional state $\Gamma^\mathrm{c}$.

  The effective equation of motion \eqref{eq.system_DC} is not manifestly in Lindblad form. In order to bring it into the Lindblad form we rewrite it as
  \begin{align*}\label{eq.DC_Lindblad}
    \dd\rho_S&=-i[H,\rho_S]+P_{ij}\left(s_i\rho_Ss_j-\frac{1}{2}(s_js_i\rho_S+\rho_Ss_js_i)\right)\\
    &\quad+\H[i\Lambda_m^Ts]\rho_S
  \end{align*}
  where
  \begin{eqnarray}
    H &=& \frac{i}{4}s^T\left(A^{-1}(\covU+i\sigma)-(\covU-i\sigma^T)A^{-T}\right)s,\\
    P &=& -\frac{1}{2}\left(A^{-1}(\covU-i\sigma)+(\covU+i\sigma^T)A^{-T}\right).\label{eq.sigma}
  \end{eqnarray}
  The individual jump operators and corresponding decay rates  are given by eigenvectors $v_i$  and eigenvalues $w_i>0$ of the matrix $P$, $\sum_iw_i\D[v_i^Ts]\rho_S$. $P$ is indeed a positive semidefinite matrix, as we show in App.~\ref{app.positivity}.

    Finally, the effective equation of motion has to be appended with an equation relating the measured photocurrent to the system observables $s_i$ after elimination of the transducer degrees of freedom (replacing Eq.~\eqref{eq.dI})
  \[
  \dd I_m = \avg{i\Lambda_m^Ts-is^\dagger\Lambda_m^\ast}\dt+\dW_m.
  \]

  (b) When the interaction and system Hamiltonians do not commute, moving to the interaction picture with respect to the system Liouvillian $\L_S$ results in a time dependent interaction.  In the simplest and most common case the system operators oscillate at a particular frequency $\pm\omega$ and the interaction Hamiltonian becomes
  \begin{align*}
  \Hint(t)& = s^T(t)r, & s(t)& = s_+e^{i\omega t}+s_-e^{-i\omega t},
  \end{align*}
  with time independent operators $(s_+)_i=(s_-)_i^\dagger$. Since in this case the signal of the system (i.e. the photons emitted by it) is now carried by sidebands it will be necessary to detune the local oscillators in the homodyne measurements, such that the canonical operators in the measurement terms become time dependent,
  $q_i = (a_ie^{-i\Delta_mt}+a_i^\dagger e^{i\Delta_mt})/\sqrt{2}$,
  $p_i = i(a_i^\dagger e^{i\Delta_mt}-a_ie^{-i\Delta_mt})/\sqrt{2}$,
  with $a_i$ being the annihilation operator for mode $i$ and $\Delta_m$ the detuning of the local oscillator from the central frequency.  The time dependence can be moved to the coefficients of the measurement operators in Eq.~\eqref{eq.sme},
  $\lambda_m = (c_m+im_m)^Tr(t) = (c_m(t)+im_m(t))^Tr$.
  Performing a coarse-graining in time (provided $\Delta_m$ is faster than any other time scale of the transducer),
  the Riccati equation \eqref{eq.Riccati} has to be replaced by
  \begin{subequations}
  \begin{eqnarray}
    \dot\covC &=& A\covC+\covC A^T+2N- \\ \nonumber
      &&-\sum_m\sum_{a\in\{\mathrm{c},\mathrm{s}\}}(\covC c_m^a-\sigma m_m^a)(\covC c_m^a-\sigma m_m^a)^T,\\
    c_m(t) &=& c_m^\mathrm{c}\cos(\Delta_mt)+c_m^\mathrm{s}\sin(\Delta_mt),\\
    m_m(t) &=& m_m^\mathrm{c}\cos(\Delta_mt)+m_m^\mathrm{s}\sin(\Delta_mt).
  \end{eqnarray}
  \end{subequations}
  Eliminating the transducer, the system density operator then obeys the equation of motion
  \begin{subequations}\label{eq.system_AC}
  \begin{eqnarray}
    \dd\rho_S &=& \L\rho_S\dt+\H[\Lambda_m]\rho_S\dW_m,\\
    \dd I_m &=& \avg{\Lambda_m+\Lambda_m^\dagger}\dt+\dW_m,
  \end{eqnarray}
  \end{subequations}
  where the deterministic part is given by
  \begin{eqnarray}
    \L\rho_S &=& \frac{1}{2}(A+i\omega)_{ij}^{-1}(\covU_{jk}[s_{+,i},[s_{-,k},\rho_S]]+\nonumber\\
      &&\qquad+i\sigma_{jk}[s_{+,i},[s_{-,k},\rho_S]_+])+\nonumber \\
      &&+\frac{1}{2}(A-i\omega)_{ij}^{-1}(\covU_{jk}[s_{-,i},[s_{+,k},\rho_S]]+\nonumber\\
      &&\qquad+i\sigma_{jk}[s_{-,i},[s_{+,k},\rho_S]_+]),
  \end{eqnarray}
  and the particular form of the measurement term depends on the choice of local oscillator detuning, for which one has to distinguish the two relevant cases $\Delta_m = \pm\omega$,
  \begin{subequations}
  \begin{eqnarray}
    \Lambda_m &=& i\Theta_m^Ts_++i\Xi_m^Ts_-,\quad \Delta_m = -\omega,\\
    \Lambda_m &=& i\Xi_m^Ts_++i\Theta_m^Ts_-,\quad \Delta_m = \omega,\\
    \Theta_m &=& (\covC-i\sigma)(Q+i\Delta_m)^{-T}c_m^++\nonumber\\
      &&+(A-i\Delta_m)^{-1}(\covC c_m^+-\sigma m_m^+),\\
    \Xi_m &=& (\covC-i\sigma)(Q-i\Delta_m)^{-T}c_m^-+\nonumber\\
      &&+(A+i\Delta_m)^{-1}(\covC c_m^--\sigma m_m^-),\\
    Q &=& A- \sum_m\sum_{a\in\{\mathrm{c,s}\}} (\covC c_m^a-\sigma m_m^a)(c_m^a)^T,
  \end{eqnarray}
  \end{subequations}
  where we denote $c_m(t) = c_m^+e^{i\Delta_mt}+c_m^-e^{-i\Delta_mt}$ and $m_m^\pm$ are defined similarly.
  One should, once again, check that each measurement term has a corresponding Lindblad term
  in the unconditional part of the dynamics.

  In the rest of this section, we present detailed derivations of the equations of motion Eqs.
  \eref{eq.system_DC}, \eref{eq.system_AC}.
  Reader interested in applications of these results may thus jump straight to Sec. \ref{sec.examples},
  where we illustrate the use of these equations on several examples concerning qubit readout in circuit QED.

\subsection{Adiabatic elimination with time-independent interaction}\label{ssec.DC}

  We start the adiabatic elimination by simply tracing out the transducer dynamics from the stochastic master equation \eref{eq.sme},
  leading to
  \begin{equation}\label{eq.rho_S}
    \dd\rho_S = \tr_T(\dd\rho) = -i\epsilon[s_i,\eta_i]\dt+2c_{mi}\mu_i\dW_m,
  \end{equation}
  where we defined
  \begin{align}\label{eq.etamu}
  \eta_i &= \tr_T(r_i\rho) & \mu_i &= \eta_i-x_i\rho_S.
  \end{align}
  In view of $\mu_i=\tr_T(r_i\rho)-\tr_T(r_i\rho_T)\tr_T\{\rho\}$ we can give a simple physical meaning to the quantities $\mu_i$: They measure the deviation of the exact state $\rho$ from the tensor product state $\rho_T\otimes\rho_S$ with respect to the first order moments of the transducer's canonical variables $r_i$.  Accordingly, for the tensor product state $\rho=\rho_T\otimes\rho_S$ we have $\mu_i=0$, and, as we will see, the $\mu_i$ are of first order in $\epsilon$.
  Next, we derive equations governing the evolution of $\eta_i$ and $\mu_i$, solve them formally to first order in $\epsilon$,
  and plug the solutions into \eq{eq.rho_S}.

    To obtain an equation for $\eta_i$, we need to evaluate
  \begin{eqnarray}\label{eq.eta}
    \dd\eta_i &=& \tr_T(r_i\dd\rho)\nonumber\\
      &=& \tr_T(r_i\L_T\rho)\dt-i\epsilon\tr_T(r_i[s_jr_j,\rho])\dt+\nonumber\\
        &&+c_{mj}\tr_T(r_i[r_j-x_j,\rho]_+)\dW_m+\nonumber\\
	&&+im_{mj}\tr_T(r_i[r_j,\rho])\dW_m\nonumber\\
      &=& A_{ij}\eta_j-\frac{i}{2}\epsilon(\covC_{ij}+2x_ix_j)[s_j,\rho_S]\dt+\nonumber\\
        &&+\frac{1}{2}\epsilon\sigma_{ij}[s_j,\rho_S]_+\dt
        +c_{mj}(U_{ij}+2x_i\mu_j)\dW_m-\nonumber\\
	&&-\sigma_{ij}m_{mj}\rho_S\dW_m.
  \end{eqnarray}
  Here we used the fact that
  the first term on the right hand side of the above equation is completely analogous to terms appearing in the equation of motion
  for the mean values of the canonical operators \eq{eq.conditional_means}, and is therefore equal to $A_{ij}\eta_j$.
  For the other deterministic part we further used
  \begin{eqnarray}
    \avg{r_ir_j} &=& \frac{1}{2}\avg{[r_i,r_j]_++[r_i,r_j]}\nonumber\\
      &=& \frac{1}{2}(\covC_{ij}+2x_ix_j+i\sigma_{ij}),
  \end{eqnarray}
  and $\rho = \rho_S\otimes\rho_T$ to 0th order in $\epsilon$.
  Finally, we defined $U_{ij} = \tr_T([r_i-x_i,r_j-x_j]_+\rho)$.
  To solve Equ.~\eqref{eq.eta}, we also need an equation of motion for $x_ix_j\rho_S$ that is valid to 0th order in $\epsilon$.
  Using the It\=o product rule, $\dd(XY) = (\dd{}X)Y+X\dd{}Y+\dd{}X\dd{}Y$, we have
  \begin{subequations}
  \begin{eqnarray}
    \dd x_i &=& \tr_S(\dd\eta_i)\nonumber \\
      &=& A_{ij}x_j\dt+\epsilon\sigma_{ij}\avg{s_j}\dt+\nonumber\\
        &&+(\covC_{ij}c_{mj}-\sigma_{ij}m_{mj})\dW_m\\
    \dd(x_i\rho_S) &=& (\dd x_i)\rho_S+x_i\dd\rho_S+\dd x_i\dd\rho_S\nonumber\\
      &=& A_{ij}x_j\rho_S\dt+\epsilon\sigma_{ij}\avg{s_j}\rho_S\dt+\nonumber\\
        &&+(\covC_{ij}c_{mj}-\sigma_{ij}m_{mj})(2c_{mk}\mu_k\dt+\rho_S\dW_m)-\nonumber\\
	\label{eq.xrho}
	&&-ix_i\epsilon[s_j,\eta_j]\dt+2x_ic_{mj}\mu_j\dW_m\\
    \dd(x_ix_j\rho_S) &=& A_{ik}x_kx_j\rho_S\dt+x_ix_k\rho_sA_{kj}^T\dt+ \nonumber\\
      &&+(\covC_{ik}c_{mk}-\sigma_{ik}m_{mk})(\covC_{jl}c_{ml}-\sigma_{jl}m_{ml})\rho_S\dt,\nonumber\\
      &=& A_{ik}x_kx_j\rho_S\dt+x_ix_k\rho_SA_{kj}^T\dt\nonumber\\
        &&+\frac{1}{2}(A_{ik}\covC_{kj}+\covC_{ik}A_{kj}^T+2N_{ij})\rho_S\dt,\label{eq.xxrho}
  \end{eqnarray}
  \end{subequations}
  where we used the Riccati equation \eref{eq.Riccati} in the last equation;
  moreover, we used $\mu_i = \eta_i-x_i\rho_S = O(\epsilon)$.
  Finally, we also dropped stochastic terms in the last equation since they would give rise to a stochastic contribution
  of second order in $\epsilon$ in the effective stochastic master equation.

  \eq{eq.xxrho} is a Lyapunov equation; generally the steady-state solution of a Lyapunov equation $AX+XA^T+B=0$ can be written as
  \begin{equation}
    X = \int_0^\infty\dt e^{At}Be^{A^Tt}.
  \end{equation}
  A straightforward calculation shows that in this case this amounts to
  \begin{equation}
    x_ix_j\rho_S = \frac{1}{2}(\covU_{ij}-\covC_{ij})\rho_S,
  \end{equation}
  which can be plugged into \eq{eq.eta}, which thus gets the form
  \begin{eqnarray}\label{eq.eta2}
    \dd\eta_i &=& A_{ij}\eta_j\dt-\frac{i}{2}\epsilon\covU_{ij}[s_j,\rho_S]\dt
        +\frac{1}{2}\epsilon\sigma_{ij}[s_j,\rho_S]_+\dt+\nonumber\\
      &&+(\covC_{ij}c_{mj}-\sigma_{ij}m_{mj})\rho_S\dW_m;
  \end{eqnarray}
  here we used $\rho=\rho_S\otimes\rho_T+O(\epsilon)$ which leads to $\tr_T([r_i,r_j]_+\rho)-2\eta_ix_j = \covC_{ij}\rho_S$.
  We formally solve this equation; a straightforward calculations leads to
  \begin{eqnarray}\label{eq.eta_sol}
    \eta_i &=& \frac{i}{2}\epsilon A_{ij}^{-1}\covU_{jk}[s_k,\rho_S]\dt
        -\frac{1}{2}\epsilon A_{ij}^{-1}\sigma_{jk}[s_k,\rho_S]_+\dt-\nonumber\\
      &&-A_{ij}^{-1}(\covC_{jk}c_{mk}-\sigma_{jk}m_{mk})\rho_S\dW_m.
  \end{eqnarray}
  We can already see that the unconditional part of the reduced equation will not depend on the conditional state, as expected;
  since $\mu_i$ enters \eq{eq.rho_S} only in the stochastic term,
  the unconditional part of \eref{eq.eta_sol} gives the only contribution to the unconditional dynamics of the system density operator
  $\rho_S$.

  We proceed similarly to obtain an equation of motion for $\mu_i$.
  Combining Eqs. \eref{eq.eta}, \eref{eq.xrho} and keeping terms to first order in $\epsilon$, we have
  \begin{eqnarray}\label{eq.mu}
    \dd\mu_i &=& \dd\eta_i-\dd(x_i\rho_S)\nonumber\\
      &=& A_{ij}\mu_j\dt-2(\covC_{ik}c_{mk}-\sigma_{ik}m_{mk})c_{mj}\mu_j\dt+\nonumber\\
        &&+\frac{1}{2}\epsilon\sigma_{ij}[s_j-\avg{s_j},\rho_S]_+\dt-\frac{i}{2}\epsilon\covC_{ij}[s_j,\rho_S]\dt+\nonumber\\
        &&+\Omega_{ij}c_{mj}\dW_m,
  \end{eqnarray}
  where $\Omega_{ij} = U_{ij}-\covC_{ij}\rho_S$. The quantities $\Omega_{ij}$ can be interpreted in a similar way as the $\mu_i$ in Eq.~\eqref{eq.etamu}: $\Omega_{ij}$ measures the deviation of the exact state $\rho$ from the tensor product state $\rho_T\otimes\rho_S$ with respect to the second order moments of the transducer's canonical variables $r_i$. The equation of motion for $\Omega_{ij}$ can be derived in a similar way as for $\mu_i$, and shows that this is a second order quantity $\Omega_{ij} = O(\epsilon^2)$. Therefore, the stochastic term can be dropped in \eq{eq.mu}, and the solution is
  \begin{equation}\label{eq.mu_sol}
    \mu_i = -\frac{1}{2}\epsilon Q_{ij}^{-1}\sigma_{jk}[s_k-\avg{s_k},\rho_S]_++\frac{i}{2}\epsilon Q_{ij}^{-1}\covC_{jk}[s_k,\rho_S],
  \end{equation}
  where $Q = A-2(\covC c_m-\sigma m_m)c_m^T$.
  Plugging the results \eref{eq.eta_sol}, \eref{eq.mu_sol} into the equation of motion of the system density operator,
  \eq{eq.rho_S}, we recover the effective equation \eref{eq.system_DC}.  To show that the resulting equation is a valid Be\-lav\-kin equation,  we need to show that each measurement channel has a corresponding decay term. This issue is addressed in  Appendix \ref{app.positivity}.


  \subsection{Oscillating system operators}\label{ssec.AC}

  When moving to the rotating frame with respect to the system Hamiltonian,
  the interaction stays time independent only for special cases.
  Generally, the system operators will become time dependent.
  To go beyond the model presented in Sec. \ref{ssec.DC}, we now consider the simplest case of time dependent operators---those
  oscillating at frequency $\pm\omega$.
  We thus write the interaction Hamiltonian as $\Hint = s^T(t)r$, where $s(t) = s_+e^{i\omega t}+s_-e^{-i\omega t}$,
  and the operators $s_\pm$ are time independent.
  Although this is not a completely general form of system-transducer coupling, together with the time-independent case,
  it can cover a large range of scenarios, including arbitrary qubit dynamics.

  Since the system operators now oscillate at frequency $\omega$, the essential part of the signal will no longer be transmitted
  in the carrier frequency of the transducer but by the sidebands instead.
  To recover this signal, we perform the measurements with local oscillators that are detuned from the standard reference frame.
  Denoting the frequency of the standard reference frame
  (corresponding, e.g., to the frequency of the laser light used for the readout) as $\omega_0$
  and the frequency of the local oscillator as $\omega_m$, we can follow the approach of Ref. \cite{Hofer2014}.
  This adjustment results in time-dependent measurement operators $\lambda_m = (c_m+im_m)^Tr(t)$, where we have
  \begin{equation}
    q_i = \frac{a_ie^{-i\Delta_mt}+a_i^\dagger e^{i\Delta_mt}}{\sqrt{2}},\quad
      p_i = i\frac{a_i^\dagger e^{i\Delta_mt}-a_ie^{-i\Delta_mt}}{\sqrt{2}}
  \end{equation}
  with $\Delta_m = \omega_0-\omega_m$.
  Alternatively, we can also rewrite the measurement operators so that the time dependence enters through the coefficients,
  $\lambda_m(t) = (c_m(t)+im_m(t))^Tr$, which will prove useful when adiabatically eliminating the transducer dynamics.
  Overall, the stochastic master equation thus takes the form
  \begin{equation}
    \dd\rho = -i\epsilon[s^T(t)r,\rho]\dt+\L_T\rho\dt+\sum_m\H[\lambda_m(t)]\rho\dW_m,
  \end{equation}
  where we explicitly write the time dependence of the interaction Hamiltonian and the measurement operators.

  Before proceeding with the elimination procedure,
  some attention has to be paid to the conditional steady state of the Gaussian system.
  Since the measurement terms are now time-dependent, the Riccati equation \eref{eq.Riccati} for this system is ill-defined.
  To circumvent this problem, we perform a rotating wave approximation in the measurement terms
  by introducing the coarse-grained Wiener increments
  \begin{subequations}
  \begin{eqnarray}
    \dW_m^\mathrm{c} &=& \int\sqrt{2}\cos(\Delta_mt)\dW,\\
    \dW_m^\mathrm{s} &=& \int\sqrt{2}\sin(\Delta_mt)\dW.
  \end{eqnarray}
  \end{subequations}
  For integration intervals long on the time scale of $\Delta_m^{-1}$ but short on all other time scales,
  this produces two independent Wiener increments, $\dW_m^a\dW_n^b = \delta_{mn}\delta_{ab}\dt$,
  $a,b = \{\mathrm{c},\mathrm{s}\}$, effectively turning every measurement into two,
  \begin{equation}
    \H[\lambda_m(t)]\rho\dW_m \to \frac{1}{\sqrt{2}}\H[\lambda_m^\mathrm{c}]\rho\dW_m^\mathrm{c}
       +\frac{1}{\sqrt{2}}\H[\lambda_m^\mathrm{s}]\rho\dW_m^\mathrm{s},
  \end{equation}
  where $\lambda_m^a = (c_m^a+im_m^a)^Tr$ and
  \begin{subequations}
  \begin{eqnarray}
    c_m(t) &=& c_m^\mathrm{c}\cos(\Delta_mt)+c_m^\mathrm{s}\sin(\Delta_mt),\\
    m_m(t) &=& m_m^\mathrm{c}\cos(\Delta_mt)+m_m^\mathrm{s}\sin(\Delta_mt).
  \end{eqnarray}
  \end{subequations}
  These measurement operators are time-independent and thus give rise to a valid Riccati equation
  \begin{eqnarray}
    \dot\covC &=& A\covC+\covC A^T+2N- \\ \nonumber
      &&-\sum_m\sum_{a\in\{\mathrm{c},\mathrm{s}\}}(\covC c_m^a-\sigma m_m^a)(\covC c_m^a-\sigma m_m^a)^T.
  \end{eqnarray}
  We treat the matrix $Q$ [\eq{eq.q}] which now also becomes time-dependent, in a similar manner; it becomes
  \begin{equation}
    Q = A- \sum_m\sum_a (\covC c_m^a-\sigma m_m^a)(c_m^a)^T.
  \end{equation}
  With these adjustments, we are now ready to adiabatically eliminate the transducer dynamics,
  and obtain an effective equation for the system.

  Since we made no assumptions about time dependence of the system operators in deriving equations of motion
  for $\rho_S$, $\eta_i$, $x_ix_j\rho$, and $\mu_i$, Eqs. \eref{eq.rho_S}, \eref{eq.eta}, \eref{eq.xxrho}, \eref{eq.mu},
  these equations are valid also in the present case.
  It is only their formal solution, where the time dependence of the system and measurement operators starts to play a role.
  The solution is, nevertheless, analogous to the time independent case, only with additional oscillation terms,
  $e^{\pm i\omega t}$.
  Solving the equations of motion for $x_ix_j\rho_S$, $\eta_i$, $\mu_i$ formally and
  performing the rotating wave approximation, keeping only stationary terms,
  a straightforward calculation recovers \eq{eq.system_AC}.
  To bring the deterministic part of this equation to Lindblad form, we can proceed similar to the time-independent case.
  Since now the system operators $s_{\pm,i}$ are non-Hermitian, we first need to express them using some Hermitian basis
  (in the case of qubits, for instance, that would be the set of the Pauli operators and the identity).
  We can then recover the Hamiltonian part and the dissipative part,
  the diagonalization of which reveals the individual decay channels.
  It then remains to show that the measurement channels are included in the decay, for which we refer to Appendix \ref{app.positivity}.


  \section{Examples}\label{sec.examples}

  In this section, we illustrate the use of the adiabatic elimination method presented in Sec. \ref{sec.elimination}
  on a few simple examples.
  The model scenarios are taken from circuit QED where thermal noise---typically not accounted for
  by other adiabatic elimination methods---can be present even in cryogenically cooled systems.
  We show that our adiabatic elimination method, which we henceforth refer to as Gaussian adiabatic elimination,
  can provide significantly increased accuracy for thermal noise at the level of few quanta.

  \begin{figure}
    \centering
    \includegraphics[width=0.9\linewidth]{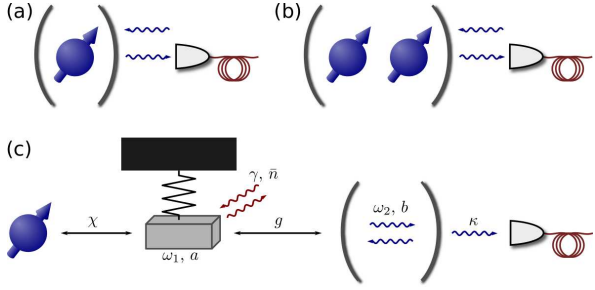}
    \caption{\label{fig.examples}
      Schematic illustrations of the setups we consider to illustrate the Gaussian adiabatic elimination method.
      In Secs. \ref{ssec.QND}, \ref{ssec.JC}, we analyze dynamics of a qubit coupled to a thermal cavity
      via dispersive and Jaynes-Cummings interaction, respectively, shown in (a).
      (b) Setup for entanglement generation by measurement as discussed in Sec. \ref{ssec.parity}.
      Here, two qubits interact dispersively with the same cavity but not with each other.
      Using the measurement record, it is then possible to postselect an entangled state of the two qubits.
      In Sec. \ref{ssec.OM_transducer}, we study a qubit coupled to a two-oscillator transducer,
      where the first oscillator couples to a thermal bath and the second oscillator is used for readout of the qubit state,
      as shown in (c).
    }
  \end{figure}

  The examples we consider are illustrated in Fig. \ref{fig.examples}.
  In Sec. \ref{ssec.QND}, we consider dispersive readout of a qubit from a cavity that is coupled to a thermal reservoir,
  see Fig. \ref{fig.examples}(a).
  We compare Gaussian adiabatic elimination with results obtained by density operator expansion
  and show that significant qualitative and quantitative improvements can be achieved with the former method.
  We extend this system in Sec. \ref{ssec.parity} where we study the effect thermal noise has on generating two-qubit entanglement
  by measurement, following the approach of Ref. \cite{Hutchison2009} [Fig. \ref{fig.examples}(b)].
  Next, we illustrate the use of Gaussian adiabatic elimination with time-dependent interaction in Sec. \ref{ssec.JC}
  where we consider a single qubit coupled to a cavity field via Jaynes-Cummings Hamiltonian.
  Finally, in Sec. \ref{ssec.OM_transducer}, we consider the system shown in Fig. \ref{fig.examples}(c)---a
  transducer consisting of two coupled oscillators, one of which is coupled to a thermal bath.
  This setup differs from all other scenarios considered here by having a different unconditional and conditional steady state of the transducer,   and we show how the Gaussian adiabatic elimination fares in this case.
  All numerical calculations in this section are done using QuTiP \cite{Johansson2012, Johansson2013};
  for more details on the particular implementations of these examples, the reader ought to refer to Ref. \cite{Cernotik2015}.


  \subsection{Single qubit QND readout}\label{ssec.QND}

    We consider the system shown in Fig.~\ref{fig.examples}(a), where a qubit couples in a QND interaction to a cavity mode whose output file is subject to continuous homodyne detection.  In such a system the cavity itself serves just as a transducer, and can be adiabatically eliminated if the cavity decay rate is sufficiently large (faster than the QND coupling).  In this case, adiabatic elimination is usually based on expanding the density operator in the Fock basis of the cavity around its vacuum state assuming no thermal excitations in the cavity \cite{Doherty1999, Hutchison2009},
  \begin{equation}
    \rho = \rho_{00}\outpr{0}{0}+\rho_{10}\outpr{1}{0}+\rho_{01}\outpr{0}{1}+\rho_{11}\outpr{1}{1}+\ldots,
  \end{equation}
  where the elements $\rho_{ij}$ are operators acting on the Hilbert space of the system, and are of the order $i+j$ in the small coupling parameter $\epsilon$.
  Expanding up to second order, the reduced state of the qubit is given by $\rho_S = \tr_T(\rho) = \rho_{00}+\rho_{11}$,
  and the elements $\rho_{00}$, $\rho_{11}$ depend on $\rho_{ij}$ with $i+j\le 2$. However, such an approach is limited to zero temperature where the cavity is essentially in vacuum. Our method allows to drop this assumption, and take thermal noise in the cavity into account in a systematic fashion.

  Before we illustrate our method on the basis of this example we note that more refined versions of adiabatic eliminations exist which employ a polaron-like transformation  \cite{Gambetta2008, Lalumiere2010}, and cover regimes of strong interactions between cavity and qubit. We believe that similar approach, i.e., using different conditional steady states for different states of the system,
  is possible also with Gaussian adiabatic elimination; such a generalization is, however, beyond the scope of the present paper.

  We start from the standard dispersive interaction with the qubit-cavity Hamiltonian of the form
  \begin{equation}
    H = \frac{\omega}{2}\sigma_z+\Delta a^\dagger a+ga^\dagger a\sigma_z+(\varepsilon^\ast a+\varepsilon a^\dagger).
  \end{equation}
  We move to the interaction picture with respect to the free qubit evolution, canceling the first term.
  The second term gives the free cavity dynamics; we choose to drive the cavity mode at the center frequency, $\Delta = 0$,
  maximizing the measurement efficiency.  The third term gives the standard dispersive interaction, and the last term describes the cavity drive.

  To obtain an interaction that is linear in the cavity quadrature operators,
  we linearize the Hamiltonian by moving to the displaced frame, $\rho\to D^\dagger(\alpha)\rho D(\alpha)$
  with $D(\alpha) = \exp(\alpha a^\dagger-\alpha^\ast a)$ being the displacement operator,
  and $\alpha = -2i\varepsilon/\kappa$, where $\kappa$ is the cavity decay rate.
  (The linearization also makes it possible to eliminate the cavity field using density operator expansion approach.)
  This procedure brings the interaction Hamiltonian to the form $g(\alpha^\ast a+\alpha a^\dagger)\sigma_z+ga^\dagger a\sigma_z$.
  If the driving field is strong enough, we can drop the second term, getting the interaction Hamiltonian
  $\Hint = \chi r_\phi\sigma_z$, where $r_\phi = \frac{1}{\sqrt{2}}(ae^{i\phi}+a^\dagger e^{-i\phi})$
  and $\chi = \sqrt{2}g|\alpha|$. The phase $\phi$ is set by the field $\varepsilon$ driving the cavity.

  Since the cavity field couples to a thermal bath, the measurement term takes the form
  $\sqrt{\frac{\kappa}{2\nbar+1}}\H[(\nbar+1)a-\nbar a^\dagger]\rho$ \cite{Wiseman2010}.
  The full dynamics of the qubit-cavity system is thus described by the equation
  \begin{eqnarray}\label{eq.QND_full}
    \dd\rho &=& -i\chi[\sigma_zr_\phi,\rho]\dt+\kappa\{(\nbar+1)\D[a]+\nbar\D[a^\dagger]\}\rho\dt+\nonumber\\
      &&+\sqrt{\frac{\kappa}{2\nbar+1}}\H[(\nbar+1)a-\nbar a^\dagger]\rho\dW.
  \end{eqnarray}
  Here, we assume that the cavity leaks only through its output port at rate $\kappa$ and the homodyne detector has unit efficiency;
  our numerical simulations indicate that additional decay has little effect on the accuracy of the adiabatic elimination methods.
  The measurement signal has the form $\dd I = \sqrt{2\kappa/(2\nbar+1)}\avg{q}\dt+\dW$.

  Following the recipe from Sec. \ref{ssec.results},
  we have for the transducer Hamiltonian $H_T = 0$,
  jump operators $j_1 = \sqrt{\kappa(\nbar+1)}\,a$, $j_2 = \sqrt{\kappa\nbar}\,a^\dagger$,
  and measurement operator $\lambda = \sqrt{\kappa/(2\nbar+1)}((\nbar+1)a-\nbar a^\dagger)$, or
  \begin{subequations}
  \begin{eqnarray}
    R &=& 0,\\
    \xi_1 &=& \sqrt{\frac{\kappa(\nbar+1)}{2}}\left(\begin{array}{c}1\\i\end{array}\right),\label{eq.xi1}\\
    \xi_2 &=& \sqrt{\frac{\kappa\nbar}{2}}\left(\begin{array}{c}1\\-i\end{array}\right),\label{eq.xi2}\\
    c &=& \sqrt{\frac{\kappa}{2(2\nbar+1)}}\left(\begin{array}{c}1\\0\end{array}\right),\\
    m &=& \sqrt{\frac{\kappa(2\nbar+1)}{2}}\left(\begin{array}{c}0\\1\end{array}\right).
  \end{eqnarray}
  \end{subequations}
  It then follows that $A = -\frac{\kappa}{2}\mathds{1}$, $N = (\nbar+\frac{1}{2})\mathds{1}$
  ($\mathds{1}$ being the identity matrix),
  and both the unconditional and conditional steady state is the thermal state $\covU = \covC = (2\nbar+1)\mathds{1}$.
  Furthermore, from the interaction Hamiltonian, we can read off $s = \chi\sigma_z(\cos\phi,-\sin\phi)^T$.
  Plugging everything into \eq{eq.system_DC}, a straightforward calculation reveals the effective equation
  \begin{subequations}\label{eq.QND_GAE}
  \begin{eqnarray}
    \dd\rho_S &=& \frac{2\chi^2}{\kappa}(2\nbar+1)\D[\sigma_z]\rho_S\dt+\\ \nonumber
      &&+\sqrt{\frac{2\chi^2}{\kappa(2\nbar+1)}}\H\left[-i(2\nbar\cos\phi+e^{-i\phi})\sigma_z\right]\rho_S\dW,\\
    \dd I &=& -\sqrt{\frac{8\chi^2}{\kappa(2\nbar+1)}}\sin\phi\avg{\sigma_z}\dt+\dW. \label{eq.QND_GAE_dI}
  \end{eqnarray}
  \end{subequations}
  Obviously, the optimal phase for an efficient readout of the qubit state is $\phi = \pi/2$, which is not surprising---this
  phase choice corresponds to an interaction of the from $\Hint = \chi\sigma_z p$ accompanied by a $q$ measurement.

  In contrast, using the density matrix expansion method, the qubit equation of motion takes the form
  \begin{subequations}\label{eq.QND_DOE}
  \begin{eqnarray}
    \dd\rho_S &=& \frac{2\chi^2}{\kappa}(2\nbar+1)\D[\sigma_z]\rho_S\dt+\nonumber\\
      &&+\sqrt{\frac{2\chi^2}{\kappa}}\H[e^{-i(\phi+\pi/2)}\sigma_z]\rho_S\dW,\\
    \dd I &=& -\sqrt{\frac{8\chi^2}{\kappa}}\sin\phi\avg{\sigma_z}\dt+\dW.
  \end{eqnarray}
  \end{subequations}
  Here we took into account the effect of thermal noise in the deterministic part (first line),
  which can be easily done using, e.g., projection operator method \cite{Gardiner2004}.
  The only difference between Eqs. \eref{eq.QND_GAE} and \eref{eq.QND_DOE} is thus in the measurement term.
  Qualitatively speaking, the density operator expansion approach overestimates the strength of the measurement
  by a factor of $\sim 1/\sqrt{\nbar}$.
  This means that for a zero temperature bath, both methods give the same results.
  In the presence of thermal excitations, however, this difference quickly starts to play a role.

  \begin{figure}
    \centering
    \includegraphics[width=\linewidth]{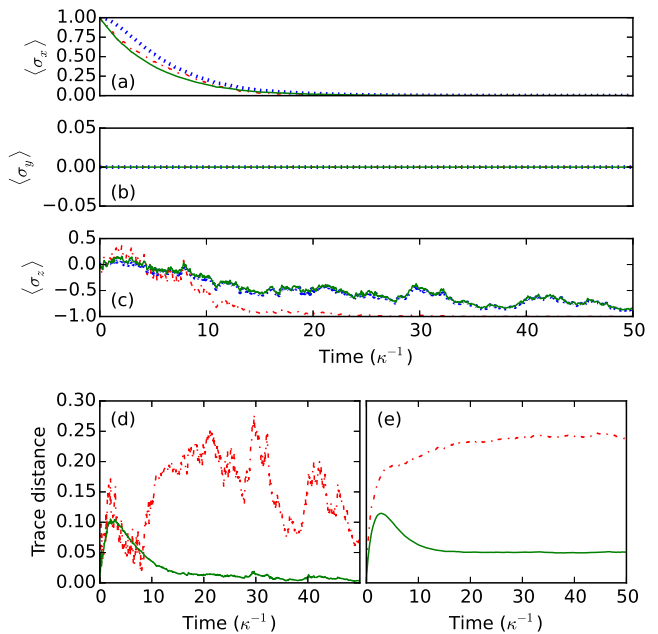}
    \caption{\label{fig.averaging}
      Illustration of determination of the average trace distance.
      Starting from a single quantum trajectory [expectation values of the Pauli matrices shown in plots (a)-(c)],
      we calculate the trace distance between the full model, \eq{eq.QND_full}, (dotted blue line)
      and the result obtained by Gaussian adiabatic elimination, \eq{eq.QND_GAE}, (full green line)
      or using the density operator expansion, \eq{eq.QND_DOE}, (dot-dashed red line).
      The resulting trace distances are shown in (d).
      We further average using 500 quantum trajectories (e) to obtain an average trace distance.
      Using time averaging on this result, we further obtain a single figure of merit that determines quality of the two approaches.
      For the results shown here with $\nbar = 2$, $\chi = 0.1\kappa$, $\phi = \pi/2$
      and initial qubit state $|\psi_0\rangle = (|0\rangle+|1\rangle)/\sqrt{2}$,
      we have the average trace distance $D\approx 0.05$ for Gaussian adiabatic elimination
      and $D\approx 0.22$ for density operator expansion (cf. Fig. \ref{fig.qnd_results}).
    }
  \end{figure}

  To quantify the difference between the full model given by \eq{eq.QND_full}
  and the effective qubit equation \eref{eq.QND_GAE} or \eref{eq.QND_DOE},
  we calculate the trace distance between the corresponding qubit states
  (we use $\rho_1$ to denote state obtained from the exact dynamics and $\rho_2$ for the approximation methods),
  $D(\rho_1,\rho_2) = \frac{1}{2}\tr|\rho_1-\rho_2|$ with $|X| = \sqrt{X^\dagger X}$.
  Since the density matrices describe the state of a single qubit, the trace distance can be expressed using the expectation values
  of the Pauli matrices $\avg{\sigma_i^j} = \tr(\rho_j\sigma_i)$ as
  \begin{equation}
    D(\rho_1,\rho_2) = \frac{1}{2}\sqrt{\sum_{i\in\{x,y,z\}}(\avg{\sigma_i^1}-\avg{\sigma_i^2})^2}.
  \end{equation}
  To obtain an average trace distance between the full model and the reduced dynamics,
  we generate a large number of quantum trajectories.
  We are thus able to study how the average trace distance changes in time;
  in addition, upon time averaging, we obtain a single figure of merit quantifying the discrepancy
  between the full and reduced dynamics;
  the averaging process is illustrated in Fig. \ref{fig.averaging}.

\begin{figure}
    \centering
    \includegraphics[width=\linewidth]{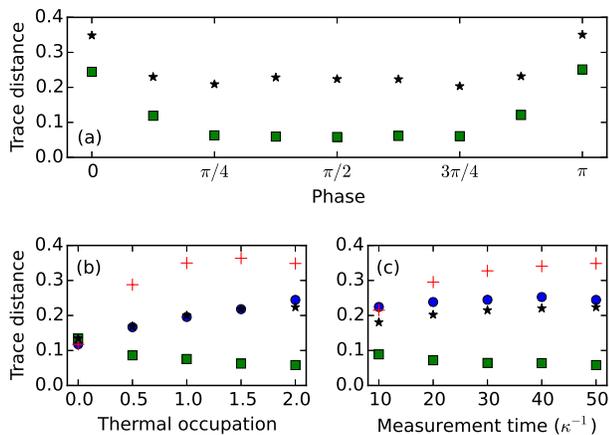}
    \caption{\label{fig.qnd_results}
      Average trace distance for Gaussian adiabatic elimination and density operator expansion as compared to the full model.
      In (a), we plot the trace distance as a function of the measurement phase
      with green squares showing results for the Gaussian adiabatic elimination and black stars for density operator expansion.
      The bottom panels show the trace distance versus thermal occupation number (b) and the overall measurement time (c)
      for two choices of phase---$\phi=\pi/2$ (green squares for Gaussian adiabatic elimination,
      black stars for density operator expansion), and $\phi = 0$ (blue circles for Gaussian adiabatic elimination,
      red crosses for density operator expansion).
      The parameters used for the simulations are $\chi = 0.1\kappa$, $\nbar = 2$ [for (a) and (c)],
      measurement time $T_m = 50$ [(a) and (b)], and initial qubit state $|\psi_0\rangle = (|0\rangle+|1\rangle)/\sqrt{2}$.
      The Fock space of the cavity field for the full model is cut off at $N_\mathrm{max} = 20$.
    }
  \end{figure}

  The results of the numerical investigations are shown in Fig. \ref{fig.qnd_results}.
  In (a), we plot the average trace distance as a function of the interaction phase $\phi$
  for Gaussian adiabatic elimination (green squares) and density operator expansion (black stars).
  We can see that both methods provide best results for $\phi = \pi/2$,
  corresponding to an interaction of the form $\Hint = \chi p\sigma_z$.
  This feature is particularly beneficial since, as discussed before,
  this phase choice is optimal for non-demolition readout of the qubit state.

  In panel (b) of Fig. \ref{fig.qnd_results}, we plot the average trace distance versus thermal occupation number.
  While the average trace distance with Gaussian adiabatic elimination (green squares for $\phi = \pi/2$, blue circles for $\phi=0$)
  eventually saturates (with the phase $\phi = 0$ this happens at $\nbar\approx 3$, which is not shown in the plot),
  the error with density operator expansion (black stars for $\phi = \pi/2$, red crosses for $\phi = 0$), as expected,
  grows with increasing temperature.
  Moreover, the Gaussian adiabatic elimination performs a factor of about 2 better than density operator expansion
  already for half a thermal excitation present in the bath;
  with the phase choice $\phi = \pi/2$, which corresponds to the optimal qubit readout,
  the difference between the two methods quickly grows.

  In Fig. \ref{fig.qnd_results}(c), we investigate how the measurement time affects the accuracy of the two methods.
  Gaussian adiabatic elimination remains unaffected by the length of the measurement ($\phi = 0$)
  or even improves with time ($\phi = \pi/2$),
  whereas accuracy of the density operator expansion method slowly deteriorates over time.
  This feature can be seen already from the time-dependence of the trace distance [cf. Fig. \ref{fig.averaging} (e)],
  where the trace distance with Gaussian adiabatic elimination reaches a maximum shortly after the begin of the evolution
  ($t\approx 5$) and then settles at a smaller steady state value,
  while the trace distance with density operator expansion continues to grow throughout the evolution.

  Finally, we note that the choice of a single initial qubit state $|\psi_0\rangle = (|0\rangle+|1\rangle)/\sqrt{2}$
  does not affect the completeness of our analysis.
  Since the evolution for eigenstates of the $\sigma_z$ operator is trivial,
  the dynamics starting from the eigenstates of $\sigma_{x,y}$ is the most interesting from the point of view of solution accuracy.
  As there is no preferred phase for the qubit, the adiabatic elimination methods perform similarly for all these states.
  We also remark that generating quantum trajectories with the approximation methods is, due to smaller size of the Hilbert space,
  about four times faster than with the full model; in systems with larger thermal noise, this effect will be even larger.
  Moreover, as the qubit dynamics happens on a slower time scale than the evolution of the cavity field,
  it is possible to use larger time steps in the numerical solution, speeding the numerics up even more.


  \subsection{Two-qubit entanglement by measurement}\label{ssec.parity}

  Extending the system presented in the previous section, we now consider two qubits dispersively coupled to a common cavity field,
  $\Hint = \sum_j\chi_j r_\phi\sigma_z^j$, where $\sigma_z^j$ acts on the $j$th qubit.
  Such a system is of particular interest as the joint measurement of the two qubits can generate entanglement between them, as discussed in  \cite{Hutchison2009}, recently realized experimentally \cite{Roch2014} in a similar scenario.
  Indeed, a straightforward generalization of \eq{eq.QND_GAE} (with $\phi = -\pi/2$) gives the effective dynamics
  \begin{subequations}\label{eq.parity}
  \begin{eqnarray}
    \dd\rho_S &=& \frac{2}{\kappa}(2\nbar+1)\D\left[\sum_j\chi_j\sigma_z^j\right]\rho_S\dt+\nonumber\\
      &&+\sqrt{\frac{2}{\kappa(2\nbar+1)}}\H\left[\sum_j\chi_j\sigma_z^j\right]\rho_S\dW,\\
    \dd I &=& \sqrt{\frac{8}{\kappa(2\nbar+1)}}\sum_j\chi_j\avg{\sigma_z^j}\dt+\dW.\label{eq.parity_signal}
  \end{eqnarray}
  \end{subequations}
  We thus get an effective measurement of the number of excitations of the two qubits.
  If we now prepare the qubits in the state $|\psi_0\rangle = \frac{1}{2}(|0\rangle+|1\rangle)\otimes(|0\rangle+|1\rangle)$,
  engineer the interactions so that $\chi_1 = \chi_2 = \chi$,
  and postselect only those trajectories with measurement $\sigma_z^1+\sigma_z^2\approx 0$,
  the two-qubit state takes the form $|\Psi_+\rangle = (|01\rangle+|10\rangle)/\sqrt{2}$
  since there is one excitation in the system but we have no information on which of the two qubits is excited.
  Moreover, this state is also a dark state of the Lindblad term $\D[\sigma_z^1+\sigma_z^2]\rho_S$
  so it is a conditional steady state of the \SME{} \eref{eq.parity}.

  We note that this approach requires post-selection and thus generates entanglement only probabilistically.
  Using the dispersive interaction in the form $\Hint = g a^\dagger a(\sigma_z^1+\sigma_z^2)$,
  one can achieve also a true parity measurement $\sigma_z^1\sigma_z^2$ where the initial state as above collapses
  either onto $|\Psi_+\rangle$ or $|\Phi_+\rangle = (|00\rangle+|11\rangle)/\sqrt{2}$,
  generating entanglement between the two qubits deterministically \cite{Lalumiere2010, Tornberg2010, Riste2013}.
  Although we believe it possible to generalize our Gaussian adiabatic elimination to include also coupling
  that is quadratic in the quadrature operators, we leave such analysis for future work
  and focus here on the probabilistic protocol only.

\begin{figure}
    \centering
    \includegraphics[width=\linewidth]{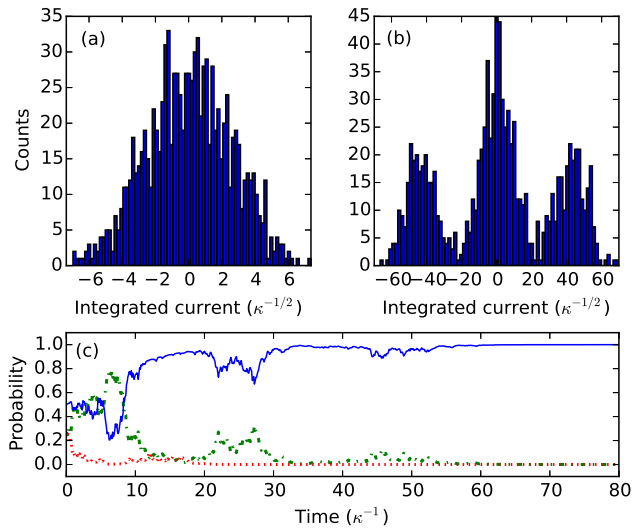}
    \caption{\label{fig.parity_hist}
      Example histograms of the integrated current at the beginning of the readout [$t = 5\kappa^{-1}$, (a)]
      and at a later time [$t = 100\kappa^{-1}$, (b)].
      In panel (c), we plot the probability of the two qubits to be in the state $|00\rangle$ (green dot-dashed line),
      $|11\rangle$ (red dotted line), and $|\Psi_+\rangle$ (blue full line).
      The simulations were run with the parameters $\chi = 0.1\kappa$, $\nbar = 0$,
      and a thousand trajectories were used for generating the histograms.
    }
  \end{figure}

  In more detail, the entanglement is generated using the following protocol:
  First, the qubits interact with the cavity mode (we assume $\chi_1 = \chi_2 = \chi$)
  and the output field is measured which is described by \eq{eq.parity}.
  After a time $T_m$, we have the integrated current
  \begin{equation}
    J(T_m) = \int_0^{T_m}\dt I(t);
  \end{equation}
  if the integrated current is close to zero, the expectation value $\avg{\sigma_z^1+\sigma_z^2}=0$
  and the qubits are in the state $|\Psi_+\rangle$.
  The whole procedure is illustrated in Fig. \ref{fig.parity_hist}.
  At an early time in the evolution [panel (a)], the distribution of the integrated current is Gaussian
  but at a later time [panel (b)] three distinct peaks form with the center one corresponding to the qubits
  in the state $|\Psi_+\rangle$.
  Quantitatively, the postselection is performed by using a threshold $\nu$ and keeping the state iff $|J|\le\nu$.
  A small threshold thus results in a pure entangled state, albeit with a small success probability;
  increasing the threshold value, in turn, results in a mixed state with reduced amount of entanglement.

\begin{figure}
    \centering
    \includegraphics[width=\linewidth]{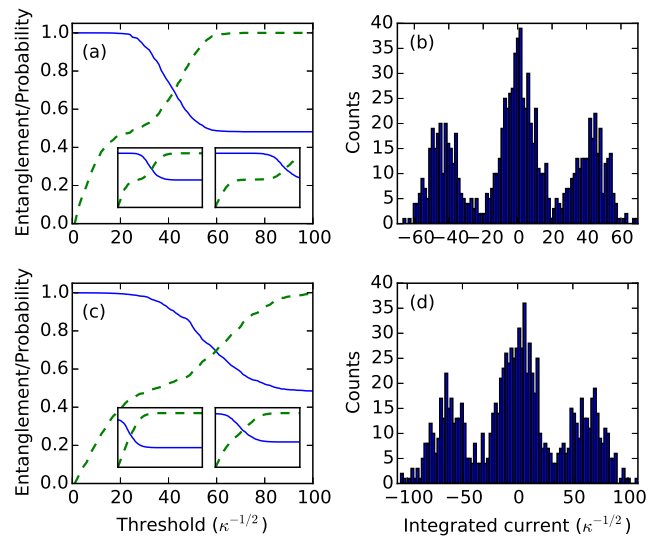}
    \caption{\label{fig.parity_results}
      Logarithmic negativity (full blue line) and success probability (dashed green line) versus the postselection threshold $\nu$
      for $\nbar = 0$ (a) and $\nbar = 2$ (c).
      The measurement time is $T_m = 100\kappa^{-1}$ in (a) and $T_m = 250\kappa^{-1}$ in (c);
      moreover, in the insets, we plot the logarithmic negativity and success probability for $T_m = 75\kappa^{-1}$ (left)
      and $T_m = 150\kappa^{-1}$ (right) for both (a), (c).
      In addition, in panels (b), (d), histograms of the integrated currents corresponding to the results in (a), (c) are shown.
      We use the coupling $\chi = 0.1\kappa$ and average over 1000 quantum trajectories.
    }
  \end{figure}

  We plot the results of the numerical simulations in Fig. \ref{fig.parity_results}.
  We analyze the logarithmic negativity \cite{Vidal2002} of the resulting postselected state (full blue line)
  and the corresponding success probability (dashed green line) for cavity coupled to a vacuum bath [panel (a)]
  and a bath with $\nbar = 2$ (c).
  Generally, in the presence of thermal photons, longer measurement times are needed to reach a maximally entangled state---we
  use the measurement times $T_m = 100\kappa^{-1}$ for zero temperature bath in panel (a),
  and $T_m = 250\kappa^{-1}$ for the results plotted in panel (c).
  This effect is due to spreading of the peaks in the integrated current with growing environment temperature,
  cf. Fig. \ref{fig.parity_results}(b), (d).
  There, one can see that the local minima between peaks are slightly less pronounced for $\nbar = 2$
  even with a measurement that is longer by a factor of 2.5.

  Our observations are further accentuated in the insets of Fig. \ref{fig.parity_results}(a), (c),
  where we plot the logarithmic negativity and success probability for $T_m = 75\kappa^{-1}$ (left inset)
  and $T_m = 150\kappa^{-1}$ (right inset).
  With thermal photons present, the logarithmic negativity does not reach unity in the limit $\nu\to 0$ for the shorter time
  while with vacuum bath, a plateau of unit entanglement starts to form.
  For longer time, we reach a large plateau of success probability of 0.5 with zero temperature,
  making it possible to generate the $|\Psi_+\rangle$ Bell state in half the cases;
  a similar plateau with the thermal bath starts to form only around $T_m = 250\kappa^{-1}$.


  \subsection{Single qubit Jaynes-Cummings readout}\label{ssec.JC}

  To illustrate the adiabatic elimination with oscillating system operators, we consider a simple example
  of a single qubit coupled to a cavity mode via Jaynes-Cummings Hamiltonian,
  $H = \frac{\omega}{2}\sigma_z+\Delta a^\dagger a+g(a\sigma_+e^{i\phi}+a^\dagger\sigma_-e^{-i\phi})$.
  Moving to the rotating frame of the qubit, this gives rise to the \SME
  \begin{eqnarray}\label{eq.JC_full}
    \dd\rho &=& -i[g(a\sigma_+e^{i(\omega t+\phi)}+a^\dagger\sigma_-e^{-i(\omega t+\phi)})+\Delta a^\dagger a,\rho]\dt+\nonumber\\
      &&+\kappa(\nbar+1)\D[a]\rho\dt+\kappa\nbar\D[a^\dagger]\rho\dt+\nonumber\\
      &&+\sqrt{\frac{\kappa}{2\nbar+1}}\H[(\nbar+1)ae^{-i\delta t}-\nbar a^\dagger e^{i\delta t}]\rho\dW.
  \end{eqnarray}
  To obtain a full model without oscillating measurement operators, we now move to the rotating frame of the cavity.  Choosing $\Delta = \omega = -\delta$, \eq{eq.JC_full} simplifies to
  \begin{subequations}\label{eq.JC_simple}
  \begin{eqnarray}
    \dd\rho &=& -ig[a\sigma_+e^{i\phi}+a^\dagger\sigma_-e^{-i\phi},\rho]\dt+\nonumber\\
      &&+\kappa(\nbar+1)\D[a]\rho\dt+\kappa\nbar\D[a^\dagger]\rho\dt+\nonumber\\
      &&+\sqrt{\frac{\kappa}{2\nbar+1}}\H[(\nbar+1)a-\nbar a^\dagger]\rho\dW,\\
    \dd I &=& \sqrt{\frac{2\kappa}{2\nbar+1}}\avg{q}\dt+\dW.
  \end{eqnarray}
  \end{subequations}
  \eq{eq.JC_simple} will be used in numerical calculations for comparison with the adiabatic elimination methods;
  it is \eq{eq.JC_full}, however, that will be used as a starting point for the elimination of the cavity dynamics.
  This choice enables us, in principle, to go beyond the scenario with $\Delta = \omega = -\delta$
  in the adiabatic approximation---using the Gaussian adiabatic elimination method,
  it is possible, for instance, to describe dynamics with measurement performed at the other sideband, $\delta = \omega$.

  The transducer dynamics is given by the Hamiltonian $H_T = \Delta a^\dagger a$,
  jump operators $j_1 = \sqrt{\kappa(\nbar+1)}\,a$, $j_2 = \sqrt{\kappa\nbar}\,a^\dagger$,
  and measurement operator $\lambda = \sqrt{\kappa/(2\nbar+1)}((\nbar+1)ae^{-i\delta t}-\nbar a^\dagger e^{i\delta t})$,
  so we have $R = \Delta\mathds{1}$, $\xi_{1,2}$ same as for the dispersive readout in Eqs. \eref{eq.xi1}, \eref{eq.xi2},
  and the measurement
  \begin{eqnarray}
    \lambda &=& \sqrt{\frac{\kappa}{2(2\nbar+1)}}\{\cos(\delta t)-i(2\nbar+1)\sin(\delta t)\}q\\
      &&+\sqrt{\frac{\kappa}{2(2\nbar+1)}}\{\sin(\delta t)+i(2\nbar+1)\cos(\delta t)\}p,\nonumber\\
    c^\mathrm{c} &=& \sqrt{\frac{\kappa}{2(2\nbar+1)}}\left(\begin{array}{c}1\\0\end{array}\right),\
    m^\mathrm{c} = \sqrt{\frac{\kappa(2\nbar+1)}{2}}\left(\begin{array}{c}0\\1\end{array}\right),\nonumber\\
    c^\mathrm{s} &=& \sqrt{\frac{\kappa}{2(2\nbar+1)}}\left(\begin{array}{c}0\\1\end{array}\right),\
    m^\mathrm{s} = \sqrt{\frac{\kappa(2\nbar+1)}{2}}\left(\begin{array}{c}0\\-1\end{array}\right),\nonumber\\
    c^+ &=& \sqrt{\frac{\kappa}{8(2\nbar+1)}}\left(\begin{array}{c}1\\-i\end{array}\right),\
    m^+ = \sqrt{\frac{\kappa(2\nbar+1)}{8}}\left(\begin{array}{c}i\\1\end{array}\right),\nonumber
  \end{eqnarray}
  and $c^- = (c^+)^\ast$, $m^- = (m^+)^\ast$.
  We thus have $A = -\frac{\kappa}{2}\mathds{1}+\Delta\sigma$, $N = (\nbar+\frac{1}{2})\mathds{1}$,
  and the cavity steady state (both unconditional and conditional) is the thermal state $\covU = \covC = (2\nbar+1)\mathds{1}$.
  Together with the system operators
  \begin{subequations}
  \begin{eqnarray}
    s_- &=& \frac{g}{\sqrt{2}}e^{-i\phi}\sigma_-\left(\begin{array}{c}1\\-i\end{array}\right),\\
    s_+ &=& \frac{g}{\sqrt{2}}e^{i\phi}\sigma_+\left(\begin{array}{c}1\\i\end{array}\right),
  \end{eqnarray}
  \end{subequations}
  and the choice of frequencies $\Delta = \omega = -\delta$, this gives the \SME
  \begin{subequations}\label{eq.JC_GAE}
  \begin{eqnarray}
    \dd\rho_S &=& \frac{4g^2}{\kappa}\{(\nbar+1)\D[\sigma_-]+\nbar\D[\sigma_+]\}\rho_S\dt+ \nonumber\\
      &&+\frac{2g}{\sqrt{\kappa(2\nbar+1)}}\H[(\nbar+1)\sigma_-e^{-i(\phi+\pi/2)}-\nonumber\\
      &&\qquad\qquad\qquad\qquad  -\nbar\sigma_+e^{i(\phi+\pi/2)}]\rho\dW,\\
    \dd I &=& \frac{2g}{\sqrt{\kappa(2\nbar+1)}}\avg{\sigma_y\cos\phi-\sigma_x\sin\phi}\dt+\dW.
  \end{eqnarray}
  \end{subequations}

  Using the density operator expansion method, together with a correction for thermal noise in the Lindblad terms,
  the qubit dynamics is described by the equation
  \begin{subequations}\label{eq.JC_DOE}
  \begin{eqnarray}
    \dd\rho_S &=& \frac{4g^2}{\kappa}\{(\nbar+1)\D[\sigma_-]+\nbar\D[\sigma_+]\}\rho_S\dt+\nonumber \\
      &&+\frac{2g}{\sqrt{\kappa}}\H[\sigma_-e^{-i(\phi+\pi/2)}]\rho_S\dW,\\
    \dd I &=& \frac{2g}{\sqrt{\kappa}}\avg{\sigma_y\cos\phi-\sigma_x\sin\phi}\dt+\dW.
  \end{eqnarray}
  \end{subequations}
  Both adiabatic elimination methods, \eq{eq.JC_GAE}, \eq{eq.JC_DOE}, give identical results for zero-temperature cavity bath.

  \begin{figure}
    \centering
    \includegraphics[width=\linewidth]{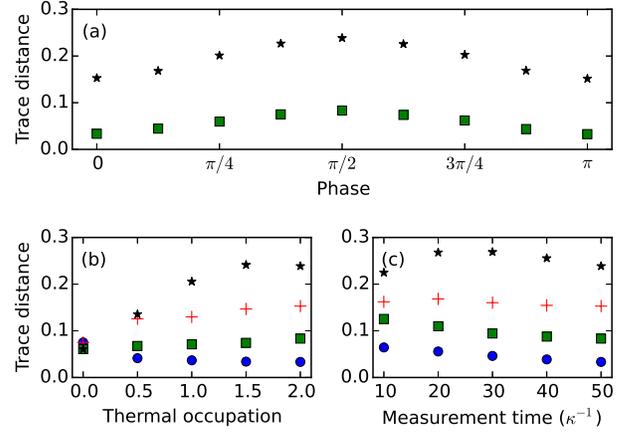}
    \caption{\label{fig.jc_results}
      (a) Average trace distance for the Gaussian adiabatic elimination (green squares)
      and density operator expansion (black stars) as a function of the interaction phase $\phi$.
      In the bottom panels, we plot the average trace distance versus thermal occupation (b) and measurement time (c)
      for the choice of phase $\phi=0$ (blue circles showing Gaussian adiabatic elimination
      and red crossess for density operator expansion) and $\phi = \pi/2$ (Gaussian adiabatic elimination shown in green squares,
      density operator expansion in black stars).
      The parameters used to run the simulations are $g = 0.1\kappa$, $\nbar = 2$ [for panels (a), (c)],
      measurement time $T_m = 50$ [for (a), (b)], and initial qubit state $|\psi_0\rangle = (|0\rangle+|1\rangle)/\sqrt{2}$.
      The cavity field for the full model has been cut off at the Fock number $N_\mathrm{max} = 20$.
    }
  \end{figure}

  The average trace distance for the Gaussian adiabatic elimination and the density operator expansion is analyzed
  in complete analogy with the dispersive readout in Fig. \ref{fig.jc_results}.
  The error is minimized for phase $\phi = 0$ (a),
  which corresponds to a $\sigma_y$ measurement, while for a $\sigma_x$ measurement (phase $\phi=\pi/2$),
  the average trace distance reaches its maximum.
  We note, however, that the Jaynes-Cummings readout is much less phase-sensitive than the dispersive readot scheme.
  Performance of the Gaussian adiabatic elimination does not depend on the thermal occupation [panel (b)]
  while the average trace distance with the density operator expansion increases as expected.
  Finally, for long measurement times [see panel (c)], the average trace distance for both methods gradually decreases
  as the measurement approaches a projective readout and the qubit approaches one of its conditional steady states
  $|0\rangle$, $|1\rangle$.


  \subsection{Two-oscillator transducer}\label{ssec.OM_transducer}

  All examples considered so far had one special property in common---the
  unconditional and conditional steady states of the transducer were equal.
  To show how Gaussian adiabatic elimination can be applied to systems where this is not the case,
  we now consider the following example, see Fig. \ref{fig.examples}(c):
  A qubit, our system of interest, couples to a harmonic oscillator by means of a quantum non-demolition interaction
  similar to the example in Sec. \ref{ssec.QND}.
  This oscillator decays into a thermal bath and, at the same time, couples to another oscillator of much higher frequency
  so its reservoir is effectively in the ground state.
  Finally, we measure the output of the second oscillator.

  The density operator of the overall system has the form
  \begin{subequations}\label{eq.two-oscillator}
  \begin{eqnarray}
    \dd\rho &=& -i[\chi\sigma_z r_\phi+\omega_1a^\dagger a+\omega_2b^\dagger b+gq_1q_2,\rho]\dt+\nonumber\\
      &&+\gamma(\nbar+1)\D[a]\rho\dt+\gamma\nbar\D[a^\dagger]\rho\dt+\nonumber\\
      &&+\kappa\D[b]\rho\dt+\sqrt{\kappa}\H[be^{i\varphi}]\rho\dW,\\
    \dd I &=& \sqrt{\kappa}\avg{be^{i\varphi}+b^\dagger e^{-i\varphi}}\dt+\dW.
  \end{eqnarray}
  \end{subequations}
  Here, $a$ describes the first (i.e., thermal) oscillator while $b$ is used for the second readout oscillator,
  and $r_\phi$ denotes a general quadrature operator of the thermal oscillator.
  Such a system can be realized by coupling superconducting qubit to a mechanical oscillator \cite{Pirkkalainen2014, Xia2014}
  and reading out the signal in the mechanical oscillator optically.
  Since the oscillator coupling has the form of standard linearized optomechanical interaction, $H_\mathrm{osc} = gq_1q_2$,
  the qubit readout is optimized by driving the readout oscillator on the red sideband, $\omega_1 = \omega_2 = \omega$.
  The readout efficiency can further be maximized by letting the qubit couple to the phase quadrature of the thermal oscillator
  and measuring the phase quadrature of the readout oscillator.
  The \SME{} then takes the form
  \begin{eqnarray}
    \dd\rho &=& -i[\chi\sigma_zp_1+\omega(a^\dagger a+b^\dagger b)+gq_1q_2,\rho]\dt+\nonumber\\
      &&+\gamma(\nbar+1)\D[a]\rho\dt+\gamma\nbar\D[a^\dagger]\rho\dt+\nonumber\\
      &&+\kappa\D[b]\rho\dt+\sqrt{\kappa}\H[ib]\rho\dW.
  \end{eqnarray}

  \begin{figure}
    \centering
    \includegraphics[width=\linewidth]{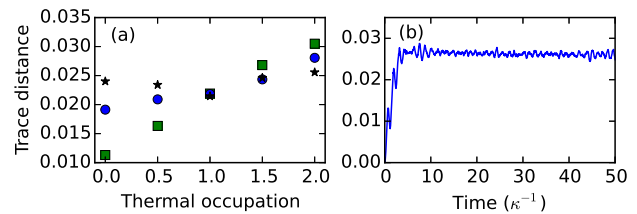}
    \caption{\label{fig.twoosc}
      (a) Average trace distance as a function of thermal occupation for three regimes:
      weak coupling ($g = 0.2\kappa$, green squares), intermediate coupling ($g = 0.5\kappa$, blue circles),
      and strong couping ($g = \kappa$, black stars).
      In (b), we show the average trace distance versus time for $\nbar = 2$, $g = \kappa$.
      Other parameters used in the simulations are $\chi = 0.2\kappa$, $\omega = 5\kappa$, $\gamma = 0.1\kappa$,
      initial qubit state $|\psi_0\rangle = (|0\rangle+|1\rangle)/\sqrt{2}$,
      and we averaged over 100 quantum trajectories.
    }
  \end{figure}

  As the transducer dynamics is more complex in this case, we perform the whole adiabatic elimination numerically,
  see Ref. \cite{Cernotik2015}.
  The results of the numerical simulations are shown in Fig. \ref{fig.twoosc}.
  In panel (a), we investigate how the bath temperature for the thermal oscillator affects the average trace distance
  in three distinct regimes---for weak ($g = 0.2\kappa$, green squares),
  intermediate ($g = 0.5\kappa$, blue circles), and strong ($g = \kappa$, black stars) coupling.
  As the strength of the coupling between the two oscillators grows, the trace distance becomes less temperature sensitive.
  In Fig. \ref{fig.twoosc}(b), we plot an example average trace distance as a function of time.
  This plot illustrates that the time dependence has features similar to simpler transducers
  considered in the previous sections---after a short initial transient time, the trace distance saturates
  and stays constant for the rest of the evolution.

  Finally, we also point out the numerical requirements for the full model and the adiabatic elimination.
  With only two thermal excitations in the heat bath, the full model needs 700 times longer time to be simulated,
  compared to the adiabatic elimination; this difference can be increased by using larger time steps for the approximate dynamics
  since the qubit evolution happens at longer time scales.
  The main limitation in our numerical analysis, however, are the memory requirements.
  With two thermal excitations (and corresponding Fock space cutoffs at 20 and 10 excitations for the thermal and readout oscillator,
  respectively), the storing of the full density matrix for the whole time evolution requires several gigabytes of working memory.
  Since the cutoff energy grows faster than linearly with increasing temperature,
  and the size of the density matrix grows quadratically with the cutoff,
  we were not able to perform reliable numerical simulations for larger bath temperatures.
  We still believe, nevertheless, that Gaussian adiabatic elimination can be used for systems
  with tens or hundreds thermal excitations present.


  \section{Conclusions}\label{sec.conclusions}

  In summary, we presented a new method of adiabatic elimination of fast degrees of freedom from stochastic quantum dynamics.
  Assuming the transducer (i.e., the system we wish to eliminate) is Gaussian,
  we can fully describe its evolution using first and second statistical moments of its canonical operators;
  moreover, the covariance matrix of the conditional state obeys deterministic Riccati equation.
  We are thus able to treat transducers coupled to thermal bath or consisting of multiple modes.
  While eliminating several modes using the approach based on density operator expansion or polaron transformation
  quickly becomes tedious, our method requires only basic linear-algebraic tools and can be easily solved numerically.

  Since the procedure we use relies on the fact that the system of interest itself has no free evolution,
  we did not present a completely general treatment---instead, we focused on the most relevant situations only.
  In the first place, we assumed that moving to the rotating frame with respect to the free Hamiltonian of the system
  leaves its interaction with the transducer time independent, corresponding, in particular, also to a quantum non-demolition interaction.
  Secondly, we considered a scenario, where the interaction has terms oscillating at the frequency $\pm\omega$.
  Adapting the method for other forms of coupling is straightforward.

  With these results, we have shown how our method can be used to simulate readout of qubits in cavity QED, as relevant e.g. to superconducting circuit QED.  Compared with the method of expanding the density operator around the vacuum state of the readout cavity,
  our method provides significantly better results already for a few thermal excitations present
  and is thus relevant to many experimental scenarios.
  We believe that further improvements can be achieved with ideas borrowed from adiabatic elimination using polaron transformation.
  There, for strong coupling between the system and the transducer,
  one has to consider different steady states of the transducer for individual states of the system
  and perform adiabatic elimination with respect to these conditional states.
  Using similar tools for our method, it should be possible to eliminate any Gaussian transducer with respect
  to several conditional steady states.
  In addition, it is possible to generalize the method for system-transducer coupling that is quadratic
  in the canonical transducer operators---one could, e.g., use such a result to analyze a full dispersive readout
  of a superconducting qubit, $\Hint = g\sigma_za^\dagger a$.

  Another field that could benefit from our results is cavity optomechanics.
  Typical frequencies of mechanical oscillations can correspond to thermal noise of a few hundred quanta
  even with cryogenic cooling.
  Such systems cannot be eliminated from stochastic master equations using present methods;
  a toy model in our last example shows how similar tasks can be achieved using our approach.
  In the future, it might be interesting to study how optomechanical systems
  used for conversion between microwave and optical fields \cite{Andrews2014, Bagci2014}
  could be used to entangle two superconducting qubits using measurement and feedback \cite{Hofer2013a}.


  \begin{acknowledgments}
    This work was funded by the European Commission (FP7-Programme) through iQUOEMS (Grant Agreement No. 323924).
    We acknowledge the support of the cluster system team at the Leibniz University Hannover in production of this work. D.V.V. acknowledges support from the Engineering and Physical Sciences Research Council (EPSRC) (EP/L020300/1).
  \end{acknowledgments}

  \appendix


  \section{Equations of motion for mean values and covariance matrix of a Gaussian system}\label{app.Gaussian}

  In this Appendix, we derive equations governing the evolution of the mean values and the covariance matrix of a Gaussian system.
  We start with the master equation
  \begin{equation}
    \dot{\rho} = -i[H,\rho]+\sum_n\D[j_n]\rho.
  \end{equation}
  Since the system is Gaussian, the Hamiltonian is quadratic in the canonical operators and we can write
  $H = \frac{1}{2}r^TRr$ (linear Hamiltonian leads to a simple displacement
  which can be treated suitably moving the origin of the phase space);
  the jump operators are linear, and we write $j_n = \xi_n^Tr = \sum_i\xi_{ni}r_i$.
  Here $r = (q_1,p_1,\ldots,q_N,p_N)^T$ is vector of the canonical operators whose commutation relations define the symplectic matrix
  \begin{equation}
    [r_i,r_j] = i\sigma_{ij},\ \sigma=\left(\begin{array}{cc}0&1\\-1&0\end{array}\right)\oplus\ldots\oplus
      \left(\begin{array}{cc} 0&1\\-1&0 \end{array}\right).
  \end{equation}
  The goal is to use the master equation to obtain equations of motion for the first and second statistical moments defined by
  \begin{equation}
    x = \avg{r} = \tr(\rho r),\qquad \Gamma_{ij} = \avg{[r_i,r_j]_+}-2x_ix_j.
  \end{equation}

  For the $i$-th mean value, we have
  \begin{eqnarray}
    \dot{x_i} &=& \tr(\dot{\rho}r_i)\nonumber \\
      &=& -i\tr\{[H,\rho]r_i\}+\sum_n\tr\{\D[j_n]\rho r_i\}\nonumber\\
      &=& -i\tr\{\rho[r_i,H]\}+\nonumber \\
        &&+\sum_n\tr\{\rho(j_n^\dagger r_ij_n-\frac{1}{2}[j_n^\dagger j_n,r_i]_+)\}.
  \end{eqnarray}
  The commutator in the first term can be rewritten as
  \begin{eqnarray}
    [r_i,H] &=& \frac{1}{2}\sum_{jk}R_{jk}[r_i,r_jr_k]\nonumber \\
      &=& \frac{1}{2}\sum_{jk}R_{jk}([r_i,r_j]r_k+r_j[r_i,r_k])\nonumber \\
      &=& \sum_{jk}\sigma_{ij}R_{jk}r_k,
  \end{eqnarray}
  where we used the fact that the Hamiltonian matrix is symmetric $R = R^T$.
  For the Lindblad terms, we have
  \begin{eqnarray}
    \tr\{\D[j_n]\rho r_i\}
      &=& \sum_{jk}\xi_{nj}\xi_{nk}^\ast\tr\{\rho(r_kr_ir_j-\frac{1}{2}[r_kr_j,r_i]_+)\}\nonumber\\
      &=& \frac{1}{2}\sum_{jk}\xi_{nj}\xi_{nk}^\ast\tr\{\rho(r_k[r_i,r_j]-[r_i,r_k]r_j)\}\nonumber \\
      &=& -\frac{i}{2}\sum_{jk}\sigma_{ij}(\xi_{nj}^\ast\xi_{nk}-\xi_{nj}\xi_{nk}^\ast)x_k.
  \end{eqnarray}
  Combining everything, we can write
  \begin{equation}
    \dot{x}_i =\sum_{jk}\sigma_{ij}R_{jk}x_k-\frac{i}{2}\sum_{njk}\sigma_{ij}(\xi_{nj}^\ast\xi_{nk}-\xi_{nj}\xi_{nk}^\ast)x_k,
  \end{equation}
  or, in the matrix form
  \begin{equation}\label{eq.dot_x}
    \dot{x} = Ax,\qquad A = \sigma R-\frac{i}{2}\sigma\sum_n(\xi_n^\dagger\xi_n-\xi_n^T\xi_n^\ast).
  \end{equation}

  For the covariance matrix, we need to evaluate
  \begin{equation}\label{eq.cov}
    \dot{\Gamma}_{ij} = \tr\{\dot{\rho}[r_i,r_j]_+\}-2(\dot{x}_ix_j+x_i\dot{x}_j).
  \end{equation}
  Similar to the previous case, we have for the coherent evolution
  \begin{eqnarray}
    [r_ir_j,H] &=& \frac{1}{2}\sum_{kl}R_{kl}(r_i[r_j,r_k]r_l+r_k[r_i,r_l]r_j+\nonumber\\
      &&+[r_i,r_k]r_jr_l+r_kr_i[r_j,r_l]),
  \end{eqnarray}
  which, combined with $[r_jr_i,H]$, gives
  \begin{equation}
    [[r_i,r_j]_+,H] = i\sum_{kl}(\sigma_{ik}R_{kl}[r_j,r_l]_+-[r_i,r_l]_+R_{lk}\sigma_{kj}).
  \end{equation}
  For the decay terms, we have
  \begin{eqnarray}
    \E[j_n]r_ir_j &=& j_n^\dagger r_ir_jj_n-\frac{1}{2}[j_n^\dagger j_n,r_ir_j]_+ \nonumber\\
      &=& \frac{1}{2}\sum_{kl}\xi_{nk}^\ast\xi_{nl}([r_k,r_ir_j]r_l+r_k[r_ir_j,r_l]) \nonumber \\
      &=& \frac{i}{2}\sum_{kl}\xi_{nk}^\ast\xi_{nl}(\sigma_{jl}r_kr_i+\sigma_{il}r_kr_j-\nonumber\\
        &&-\sigma_{jk}r_ir_l-\sigma_{ik}r_jr_l),
  \end{eqnarray}
  where we used
  \begin{eqnarray}
    [r_ir_j,r_k] &=& r_i[r_j,r_k]+[r_i,r_k]r_j \nonumber\\
      &=& i\sigma_{jk}r_i+i\sigma_{ik}r_j.
  \end{eqnarray}
  Combined with $\E[j_n]r_jr_i$ and summed over $n$, this expression gives
  \begin{eqnarray}
    \sum_n\E[j_n][r_i,r_j]_+ &=&
      \sum_{kl}\{\beta_{kl}(\sigma_{jk}[r_i,r_l]_+
      +\sigma_{ik}[r_j,r_l]_+)-\nonumber\\
      &&-2\alpha_{kl}\sigma_{jk}\sigma_{li}\},
  \end{eqnarray}
  where
  \begin{subequations}
  \begin{eqnarray}
    \alpha_{kl} &=& \frac{1}{2}\sum_n(\xi_{nk}^\ast\xi_{nl}+\xi_{nk}\xi_{nl}^\ast)\\
    \beta_{kl} &=& \frac{i}{2}\sum_n(\xi_{nk}\xi_{nl}^\ast-\xi_{nk}^\ast\xi_{nl}).
  \end{eqnarray}
  \end{subequations}
  Plugging everything into \eq{eq.cov} and using $\dot{x}_ix_j = \sum_kA_{ik}x_kx_j$,
  and writing the resulting expression in matrix form, we get the Lyapunov equation
  \begin{subequations}
  \begin{eqnarray}
    \dot{\Gamma} &=& A\Gamma+\Gamma A^T+2N,\\
    N &=& \frac{1}{2}\sigma\sum_n(\xi_n^\dagger\xi_n+\xi_n^T\xi_n^\ast)\sigma^T.
  \end{eqnarray}
  \end{subequations}

  When the dynamics is described by the conditional master equation
  \begin{equation}
    \dd\rho = -i[H,\rho]\dt+\sum_n\D[j_n]\rho\dt+\sum_m\H[\lambda_m]\rho\dW_m,
  \end{equation}
  we also need to evaluate the contributions from the measurement terms $\H[\lambda_m]\rho$.
  We start by splitting the measurement operator in its Hermitian and anti-Hermitian part,
  $\lambda_m = (c_m+im_m)^Tr = \sum_k(c_{mk}+im_{mk})r_k$, so we can write
  \begin{equation}
    \H[\lambda_m]\rho = [c^T(r-x),\rho]_++i[m^Tr,\rho].
  \end{equation}

  For the mean values, this gives the contribution
  \begin{eqnarray}
    \dd x_i &=& \tr(\H[\lambda_m]\rho)\dW_m\nonumber \\
      &=& \sum_k\tr\{c_{mk}\rho([r_i,r_k]_+-2x_kr_i\rho)+\nonumber\\
        &&+im_{mk}\rho[r_i,r_k]\}\dW_m\nonumber\\
      &=& \sum_k(\Gamma_{ik}c_{mk}-i\sigma_{ik}m_{mk})\dW_m.
  \end{eqnarray}
  The mean values thus obey the equation
  \begin{equation}\label{eq.dxAPP}
    \dd x = Ax\dt+\sum_m(\Gamma c_m-\sigma m_m)\dW_m.
  \end{equation}
  For the covariance matrix, we need to evaluate
  \begin{eqnarray}
    \dd\Gamma_{ij} &=& \tr([r_i,r_j]_+\H[\lambda_m]\rho)\dW_m-\nonumber \\
      &&-2\{(\dd{}x_i)x_j+x_i\dd{}x_j+\dd x_i\dd x_j\},
  \end{eqnarray}
  where we used the It\=o rule $\dd(XY) = (\dd X)Y+X\dd Y+\dd X\dd Y$ for the contribution of the mean values. In the following we concentrate on the stochastic contribution in the increments $\dd{}x_j$ (second term on the r.h.s. of Eq.~\eqref{eq.dxAPP}) as the deterministic contribution is trivial.
  We start by considering
  \begin{eqnarray}
    \tr([r_i,r_j]_+\dd\rho) &=& \sum_kc_{mk}\tr\{\rho([r_k,[r_i,r_j]_+]_+-\nonumber\\
      &&-2x_k[r_i,r_j]_+)\}+\nonumber\\
      &&+i\sum_km_{mk}\tr\{\rho[r_ir_j+r_jr_i,r_k]\}\nonumber\\
      &=& 2\sum_kc_{mk}(\Gamma_{ik}x_j+\Gamma_{jk}x_i)-\nonumber\\
      &&-2\sum_km_{mk}(x_i\sigma_{jk}+x_j\sigma_{ik}).
  \end{eqnarray}
  In the first sum on the right hand side, we used the fact
  \begin{equation}
    \langle[r_i,[r_j,r_k]_+]_+\rangle = 2(\Gamma_{ij}x_k+\Gamma_{jk}x_i+\Gamma_{ki}x_j+2x_ix_jx_k).
  \end{equation}
  This can be seen by comparing the third derivative of the characteristic function from the definition $\chi(r) = \tr\{D(r)\rho\}$
  (here $D(r)$ is the displacement operator) with a general Gaussian characteristic function
  \begin{equation}
    \chi(r) = \exp\{-ir^T\sigma x-\frac{1}{4}r^T\sigma^T\Gamma\sigma r\}.
  \end{equation}

  We further use
  \begin{eqnarray}
    &&(\dd x_i)x_j+x_i\dd x_j+\dd x_i\dd x_j = \nonumber\\
    &&\quad =\sum_kx_i(\Gamma_{jk}c_{mk}-\sigma_{jk}m_{mk})\dW_m+\nonumber\\
    &&\qquad +\sum_kx_j(\Gamma_{ik}c_{mk}-\sigma_{ik}m_{mk})\dW_m+\\
    &&\qquad +\sum_{k,l}(\Gamma_{ik}c_{mk}-\sigma_{ik}m_{mk})(\Gamma_{jl}c_{ml}-\sigma_{jl}m_{ml})\dt.\nonumber
  \end{eqnarray}
  Combining everything, the stochastic contributions to the covariance matrix cancel out,
  and we are left with the term
  \[
    (\Gamma c_m-\sigma m_m)(\Gamma c_m-\sigma m_m)^T.
  \]
  The dynamics of the covariance matrix is thus given by the Riccati equation
  \begin{equation}
    \dot{\Gamma} = A\Gamma+\Gamma A^T+2N-2\sum_m(\Gamma c_m-\sigma m_m)(\Gamma c_m-\sigma m_m)^T.
  \end{equation}


  \section{Positive-semidefiniteness of decay after subtracting the measurement channels}\label{app.positivity}

We start by the observation that the overall decay $\Sigma$ is positive.
  Using its definition, \eq{eq.sigma}, together with the Lyapunov equation \eref{eq.Lyapunov},
  and the definitions \eref{eq.defs}, we can write
  \begin{eqnarray}
   P &=& A^{-1}\left(N+\frac{i}{2}(A\sigma-\sigma^TA^T)\right)A^{-T}\nonumber \\
    &=& A^{-1}\sigma\sum_i\xi_i^T\xi_i^\ast\sigma^TA^{-T}\geq 0.
  \end{eqnarray}

  Next, we discuss the question whether the effective stochastic master equations in Eqs.~\eqref{eq.system_DC} and \eqref{eq.system_AC} present valid Belavkin equation, that is, whether they generate completely positive maps.  In order for this to be true each measurement channel has to have a corresponding decay process. Quantitatively, matrix $P$ in Eq.~\eqref{eq.sigma} describing all decay terms needs to be larger than the matrix $\sum_m\Lambda_m\Lambda_m^\dagger$ characterizing all measurement channels. In other words
  \[
  P' = P-\sum_m\Lambda_m\Lambda_m^\dagger
  \]
  has to be positive semidefinite.  We did not prove this statement in the general case, but checked it for all of the cases treated in Sec.~\ref{sec.examples}.

  \bibliography{AdiabaticElimination}

\end{document}